\documentclass{ifacconf}

\usepackage{graphicx}      
\usepackage{natbib}        
\usepackage[overload]{empheq}
\usepackage{amsmath,amsfonts}
\usepackage{mathtools}
\usepackage{bm}
\usepackage{xcolor}
\usepackage{tikz}
\usepackage{subcaption}
\usepackage{stmaryrd}
\usepackage{bm}
\newcommand{\kron}{\otimes}
\newcommand{\kr}{\odot}

\newenvironment{review}{\color{black}}{}
\newcommand{\rev}[1]{\begin{review}#1\end{review}}
\begin{document}
\begin{frontmatter}

\title{Decoupling multivariate functions using a non-parametric Filtered CPD approach\thanksref{footnoteinfo}} 

\thanks[footnoteinfo]{This work was supported by the Flemish fund for scientific research FWO under license number G0068.18N.}

\author[First]{Jan Decuyper} 
\author[Second]{Koen Tiels} 
\author[Third]{Siep Weiland}
\author[First,Third]{Johan Schoukens}

\address[First]{Department of Engineering Technology, Vrije Universiteit Brussel, Pleinlaan 2, 1050 Brussels, Belgium (e-mail: jan.decuyper@vub.be, johan.schoukens@vub.be).}
\address[Third]{Department of Electrical Engineering, Eindhoven University of Technology, Eindhoven, The Netherlands (e-mail: s.weiland@tue.nl)}
\address[Second]{Department of Mechanical Engineering, Eindhoven University of Technology, Eindhoven, The Netherlands (e-mail: k.tiels@tue.nl)}

\begin{abstract}                
Black-box model structures are dominated by large multivariate functions. Usually a generic basis function expansion is used, e.g.\ a polynomial basis, and the parameters of the function are tuned given the data. This is a pragmatic and often necessary step considering the black-box nature of the problem. However, having identified a suitable function, there is no need to stick to the original basis. So-called decoupling techniques aim at translating multivariate functions into an alternative basis, thereby both reducing the number of parameters and retrieving underlying structure. In this work a filtered canonical polyadic decomposition (CPD) is introduced. It is a non-parametric method which is able to retrieve decoupled functions even when facing non-unique decompositions. Tackling this obstacle paves the way for a large number of modelling applications.
\end{abstract}


\end{frontmatter}

\section{Introduction}
An essential observation on black-box model structures is that they often consist of a large number of (nonlinear) multivariate functions. 
The black-box nature of the problem dictates that the underlying complexity is unknown. It is therefore a very pragmatic and useful idea to construct a basis function expansion and tune the paramaters accordingly. Notice that this is a general practice both for static functions, e.g.\ multivariate polynomials, as well as for dynamic models, e.g.\ a \rev{Nonlinear AutoRegressive with eXogenous inputs (NARX)} model where a nonlinear multivariate function maps delayed input and output samples onto the present output \citep{billings2013}. It is in both cases necessary to include `sufficient' terms in order to be able to count on the approximation properties of the model structure. This may lead to an excessive number of parameters. For purposes of analysis there is no need to stick to the particular parametrisation. 
By changing the model parametrisation, one may achieve decoupling properties, simplify the parametrisation, and/or enable further analysis.

These insights have led to the development of so-called decoupling techniques. Given a generic multivariate nonlinear function
\begin{equation}
\label{e:1}
\boldsymbol{q} = \boldsymbol{f}(\boldsymbol{p})
\end{equation}
with $\boldsymbol{q} \in \mathbb{R}^n$ and $\boldsymbol{p} \in \mathbb{R}^m$, the prime objective of a decoupling technique is to introduce an appropriate linear transformation of $\boldsymbol{p}$, denoted $\boldsymbol{V}$, such that the nonlinear relationship may be described by univariate functions in this alternative basis. The decoupled function is then of the following form
\begin{equation}
\label{e:2}
\boldsymbol{f}(\boldsymbol{p}) = \boldsymbol{W} \boldsymbol{g}(\boldsymbol{V}^{\text{T}} \boldsymbol{p})
\end{equation}
where in the fully decoupled case, the $i$th function is $g_i(z_i)$ with $z_i = \boldsymbol{v}_i^{\text{T}} \boldsymbol{p}$, emphasising that all functions are strictly univariate. The number of allowed univariate functions, denoted $r$, is crucial since it will determine whether the implied equivalence of \eqref{e:2} exists. A second linear transformation $\boldsymbol{W}$, maps the function back onto the outputs. The linear transformations then have the following dimensions: $\boldsymbol{V} \in \mathbb{R}^{m \times r}$ and $\boldsymbol{W} \in \mathbb{R}^{n \times r}$.

\begin{figure}
\begin{center}
\includegraphics[width=0.45\textwidth]{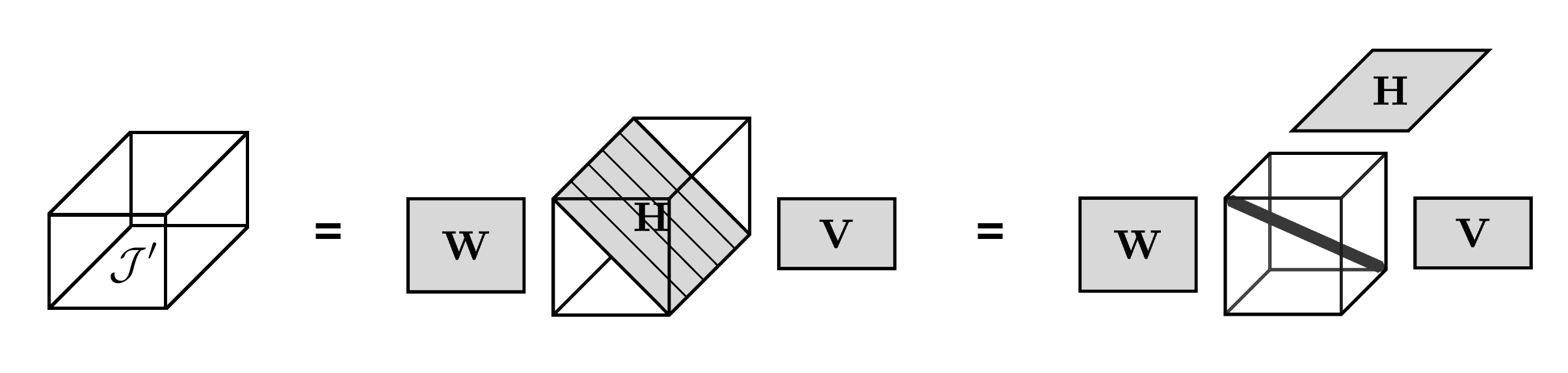}
\caption{Centre: a collection of evaluations of the Jacobian of the decoupled function (\eqref{e:2}), stacked in the third dimension. Left: the corresponding third order tensor, Right: extracting the central diagonal plane reveals a diagonal tensor decomposition.}
\label{f:1}
\end{center}
\end{figure}

It follows from \eqref{e:2} that the Jacobians of both sides of the equation need to match. This was exploited by \cite{dreesen2014} when proposing a decoupling algorithm. The idea relies on the tri-linear form which arises when computing the Jacobian of the right hand side. Denoting the left and right hand side Jacobian by $\boldsymbol{J}$ and $\boldsymbol{J}'$, respectively, we have that
\begin{equation}
\label{e:3}
\boldsymbol{J}' = \boldsymbol{W} \operatorname{diag}\left(\begin{bmatrix} h_1(z_1) \quad \cdots \quad h_r(z_r) \end{bmatrix}\right) \boldsymbol{V}^{\text{T}}
\end{equation}
in which case $h_i(z_i) \coloneqq \frac{dg_i(z_i)}{dz_i}$ represents the derivative of the univariate function $g_i(z_i)$ with respect to its argument. Evaluating this Jacobian in $N$ operating points, i.e.\ for $\{\boldsymbol{p}[1], \cdots, \boldsymbol{p}[N]\}$, and collecting the evaluations in a third dimension, expands the data object into in a three-way array $\mathcal{J}' \in \mathbb{R}^{n \times m \times N}$. The diagonal plane is denoted by $\boldsymbol{H} \in \mathbb{R}^{N \times r}$. This is illustrated graphically in Fig.~\ref{f:1}. Given the diagonal form, the collection of Jacobians may be written as a sum of $r$ outer products (or rank-one tensors). Elementwise we have that,
\begin{equation}
\label{e:outer}
j'_{jk\ell} = \sum_{i=1}^r w_{ji}~v_{ki}~h_{\ell i},
\end{equation}
is the $(j,k,\ell)$th entry of $\mathcal{J}'$.  A sum of rank-one terms defines a diagonal tensor decomposition \citep{kolda2009}. 
The latter is illustrated on the right in Fig.~\ref{f:1}. 
The decomposed tensor may be written in shorthand notation
\begin{equation}
\mathcal{J}' = \llbracket \boldsymbol{W}, \boldsymbol{V}, \boldsymbol{H} \rrbracket
\end{equation}

\begin{figure}
\begin{center}
\includegraphics[width=0.38\textwidth]{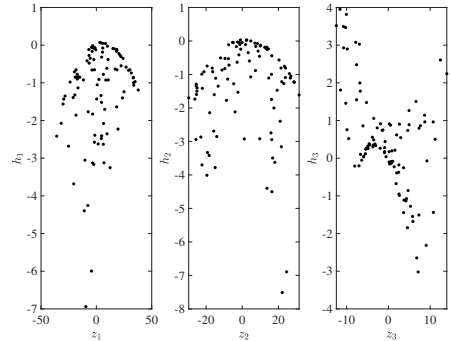}
\caption{Factor $\boldsymbol{H}$ of a non-unique tensor decomposition. The obtained solution does not correspond to the functions $h_i(z_i)$.}
\label{f:2}
\end{center}
\end{figure}

Noticing that the right hand side of Fig.~\ref{f:1} corresponds to a diagonal tensor decomposition, and recollecting that in order for the equality of Eq~\eqref{e:2} to hold, the Jacobians need to match, forms the basis of the decoupling algorithm of \cite{dreesen2014}. The method can be summarised in three steps:
\begin{enumerate}
\item Collect the Jacobian matrices of the known coupled function, $\boldsymbol{J}$, and stack them into a three-way array, i.e.\ the Jacobian tensor $\mathcal{J}  \in \mathbb{R}^{n \times m \times N}$.
\item Compute a diagonal tensor decomposition of $\mathcal{J}$, factoring it into $\{\boldsymbol{W}, \boldsymbol{V}, \boldsymbol{H}\}$. This leads to the approximation ${\mathcal{J} \approx \mathcal{J}'}$ and thus ${\mathcal{J} \approx  \llbracket \boldsymbol{W}, \boldsymbol{V},\boldsymbol{H} \rrbracket}$, where the accuracy of the approximation relies on $r$. \rev{Often a canonical polyadic decomposition or CPD is used. In that case $r$ is equal to the lowest number for which the decomposition is exact (with respect to \eqref{e:5}).}
\end{enumerate}
 According to \eqref{e:3}, the columns of $\boldsymbol{H}$ store a collection of evaluations of the functions $h_i(z_i)$, i.e.\ the derivative of $g_i(z_i)$.
\begin{enumerate}
\setcounter{enumi}{2}
\item A final step therefore parametrises the columns $\boldsymbol{h}_i$, from which the parametric functions $g_i$ may be obtained through integration.
\end{enumerate}
The method of \cite{dreesen2014} is compelling for the reason that, irrespective of the nature of the coupled function, or the size of the function in terms of $m$ and $n$, the procedure boils down to solving a third order tensor decomposition. It is, however, also subtle since it relies on two intrinsic assumptions:
\begin{enumerate}
\item The equality in \eqref{e:2}, which relies on the user-chosen value $r$.
\item The uniqueness of the tensor decomposition into its factors, $\{ \boldsymbol{W}, \boldsymbol{V},\boldsymbol{H}\}$, in order for the elements of $\boldsymbol{h}_i$ to truly correspond to evaluations of $h_i(z_i)$.
\end{enumerate}
It turns out that, untreated, both assumptions impose strong limitations on the applicability of the method.  Fig.~\ref{f:2} illustrates a potential solution of $\boldsymbol{H}$ in the case of a non-unique decomposition. \rev{The figure corresponds to the toy problem of Section \ref{s:toy} in which a third order polynomial expressed in the standard monomial basis is decoupled into three univariate branches. Although an exact solution exists, it is clear that the decomposition does not convey information on the univariate functions.} In this work, a new method is presented, leveraging the powerful idea of decomposing the Jacobian tensor while tackling the challenges of approximate decoupling and non-unique decompositions. 

\section{Problem statement}
\label{s:problem}

The objective is to have a procedure which returns $\boldsymbol{W}$, $\boldsymbol{V}$, and $\boldsymbol{g}$, together forming a decoupled function, which approximates a given arbitrary multivariate function $\boldsymbol{f}(\boldsymbol{p})$ such that for $N$ selected operating points
\begin{equation}
\label{e:4}
\underset{\boldsymbol{W},\boldsymbol{V},\boldsymbol{g}}{\operatorname{arg~min}} \frac{1}{N} \sum_{k=1}^N\lVert \boldsymbol{f}(\boldsymbol{p}[k])  - \boldsymbol{W} \boldsymbol{g}(\boldsymbol{V}^{\text{T}} \boldsymbol{p}[k]) \rVert_2^2,
\end{equation}
thereby loosening the equality of \eqref{e:2}. In practice the search space of $\boldsymbol{g}$ is limited to a family of functions, e.g.\ univariate polynomials.

Similar to \cite{dreesen2014} we will allow that the nonlinear optimisation problem of \eqref{e:4} (at the function-level), is replaced by a tri-linear\footnote{The tri-linearity may be observed from \eqref{e:outer}.} optimisation problem (at the Jacobian level)
\begin{equation}
\label{e:5}
\underset{\boldsymbol{W},\boldsymbol{V},\boldsymbol{H}}{\operatorname{arg~min}}~\lVert \mathcal{J}  - \llbracket \boldsymbol{W}, \boldsymbol{V}, \boldsymbol{H} \rrbracket \rVert_F^2,
\end{equation}
with the additional requirements that, both for
\begin{itemize}
\item non-exact decompositions, i.e.\ for values of $r$ for which $ \lVert \mathcal{J}  - \llbracket \boldsymbol{W}, \boldsymbol{V}, \boldsymbol{H} \rrbracket \rVert_F^2 \ne 0$,
\item as well as for non-unique decompositions,
\end{itemize}
the elements stored in the columns of $\boldsymbol{H}$ should remain sufficiently smooth to allow for an accurate parameterisation, using an appropriate basis expansion. Promoting smoothness will prevent that non-meaningful solutions, similar to Fig.~\ref{f:2} are obtained. Smoothness is, however, not an objective that is to be quantified in itself. The level of required smoothness goes hand in hand with the choice of basis expansion, e.g.\ the degree of the polynomial basis. What will be quantified and monitored is the accuracy of the function approximation, for which a metric is introduced in \eqref{e:rel_er}, i.e.\ a normalised version of \eqref{e:4}.

Alternative methods that aim at ensuring unique decompositions rely either on introducing information of the Hessian, or on introducing explicit polynomial constraints on the unknown $h_i$ \citep{karami2021,dreesen2018}. In this work a first-order approach is applied, avoiding the prohibitively expensive computation of the Hessian, while resorting to non-parametric methods to ensure smoothness, hence keeping the decomposition independent of any choice of basis expansion.

\section{Non-parametric Filtered-CPD}
\label{s:FCPD}

The idea revolves around introducing non-parametric finite difference filters into the decomposition routine of the Jacobian tensor. A finite difference filter is an operation which approximates the derivative operator. For each $i$, consider $N$ evaluation points and evaluations $\{z_{i}{[k]},g_{i}{[k]}\}_{k=1}^N$, such that $\boldsymbol{g}_i \coloneqq (g_i[k] \mid k=1,\ldots, N),$ and $\boldsymbol{z}_i \coloneqq(z_i[k] \mid k=1,\ldots, N)$. Given that ${h_i(z_i) \coloneqq \frac{dg_i(z_i)}{dz_i}}$, we can compute a finite difference approximation, ${\boldsymbol{h}_i \in \mathbb{R}^N}$, by applying a finite difference filter to the vector of evaluations $\boldsymbol{g}_i$,

\begin{equation}
\label{e:hg}
\boldsymbol{h}_i = \boldsymbol{S}_i^{-1} \boldsymbol{D}_i \boldsymbol{S}_i \boldsymbol{g}_i
\end{equation}
where $\boldsymbol{S}_i$ is a sorting matrix, arranging the elements of $\boldsymbol{g}_i$ in ascending order of $z_i$, and $\boldsymbol{D}_i$ embodies the finite difference operator. Should $\boldsymbol{z}_i$ be equidistantly spaced, with a spacing of $\delta_{z_i}$, $\boldsymbol{D}_i$ could for instance be defined by the Toeplitz matrix
\begin{equation}
\label{e:D}
\boldsymbol{D}_i = \frac{1}{\delta_{z_i}} \left[ \begin{array}{ccccc} 1 & -1 & 0 &  \cdots & 0 \\ 0 & 1 & -1 & \ddots & \vdots \\ \vdots & \ddots & \ddots & \ddots & 0\\0 & \dots & 0 & 1 & -1 \\ 0 & \cdots & 0 & 0 & 1 \end{array} \right]
\end{equation}
where this particular example is known as a 2-point forward differencing scheme (or a right-derivative approximation). In this context, the term \emph{finite difference filter}, is used to refer to the sequence of operations
\begin{equation}
\boldsymbol{F} \coloneqq \boldsymbol{S}^{-1} \boldsymbol{D} \boldsymbol{S}
\end{equation}
In practice, the elements of $\boldsymbol{z}_i$, will not be equidistant. A generalisation of $\boldsymbol{D}$ is provided in Appendix \ref{a:1}. Introducing finite difference filters will prove beneficial in two ways:
\begin{itemize}
\item It will allow a decomposition of $\mathcal{J}$ into $\{ \boldsymbol{W}, \boldsymbol{V},\boldsymbol{G}\}$, where $\boldsymbol{G}$ stores evaluations of the functions $g_i$, hence removing the need for an integration step (Section \ref{ss:G}).
\item It will enable to include a smoothness objective in the decomposition (Sections \ref{ss:explicit}) \citep{decuyper2019}.
\end{itemize}

\subsection{A decomposition into $\boldsymbol{W}$,$\boldsymbol{V}$, and $\boldsymbol{G}$}
\label{ss:G}
Exploiting the relationship of \eqref{e:hg} we may propose an alternative factorisation of the Jacobian tensor where the third factor contains $\boldsymbol{G}$, i.e\ a matrix holding evaluations of the \rev{(to be determined)} univariate functions $g_i(z_i)$ along its columns. This is in contrast to the factor $\boldsymbol{H}$ which stores derivative information. The sought after decomposition is then of the form
\begin{equation}
\label{e:G}
\mathcal{J} \approx \llbracket \boldsymbol{W}, \boldsymbol{V}, \mathcal{F}(\boldsymbol{V}) \circ \boldsymbol{G} \rrbracket,~\text{with}
\end{equation}
\begin{equation}
\label{e:H}
\boldsymbol{H}\color{black} \coloneqq \color{black} \mathcal{F}(\boldsymbol{V}) \circ \boldsymbol{G}
\end{equation}
The operation `$\circ$' symbolises the columnwise filtering of \eqref{e:hg}, \rev{i.e.\ computing the finite difference approximation of each branch and storing the results in the columns of $\boldsymbol{H}$}. For convenience, the finite difference filters, corresponding to each column, are stored in a three-way array $\mathcal{F} \in \mathbb{R}^{N \times N \times r}$ such that
\begin{equation} 
\mathcal{F}_{[:,:,i]}(\boldsymbol{V}) \coloneqq \boldsymbol{F}_i = \boldsymbol{S}_i^{-1} \boldsymbol{D}_i \boldsymbol{S}_i
\end{equation}
Notice the explicit dependence on the factor $\boldsymbol{V}$. The factor $\boldsymbol{V}$ defines the axes on the basis of which the finite difference is computed. Both the sorting matrix $\boldsymbol{S}_i$, and the finite difference operator $\boldsymbol{D}_i$, therefore depend on $\boldsymbol{V}$ (see Appendix \ref{a:1}). \eqref{e:H} may alternatively be written in vectorised form
\begin{equation}
\label{e:vec}
\operatorname{vec}\left(\boldsymbol{H}\right) = \left[\begin{array}{cccc}\boldsymbol{F}_1 & 0 & \dots & 0 \\ 0 & \boldsymbol{F}_2 & \ddots & \vdots \\ \vdots & \ddots & \ddots & 0 \\ 0 & \cdots & 0 & \boldsymbol{F}_r \end{array} \right] \operatorname{vec}\left(\boldsymbol{G}\right)
\end{equation}
where the vec operator transforms a matrix into a vector by concatenating its columns.
Given that the original decomposition $\{\boldsymbol{W},\boldsymbol{V},\boldsymbol{H}\}$ has been translated to $\{\boldsymbol{W},\boldsymbol{V},\boldsymbol{G}\}$, also the smoothness requirement, which was formulated in Section \ref{s:problem}, is transferred onto $\boldsymbol{G}$. 

\subsection{Adding a smoothness objective}
\label{ss:explicit}

Smoothness and finite differences are intimately linked, e.g.\ abrupt local changes will be emphasised by a finite difference filter. Appendix \ref{a:1} illustrates that multiple finite difference filters, which locally act on a different window of points, may be constructed. Given that they all approximate the first order derivative, differences amongst their results convey information on the local smoothness of the function.

\begin{figure*}
\begin{center}
\begin{subfigure}{0.3\textwidth}
\begin{center}
\includegraphics[width=0.92\textwidth]{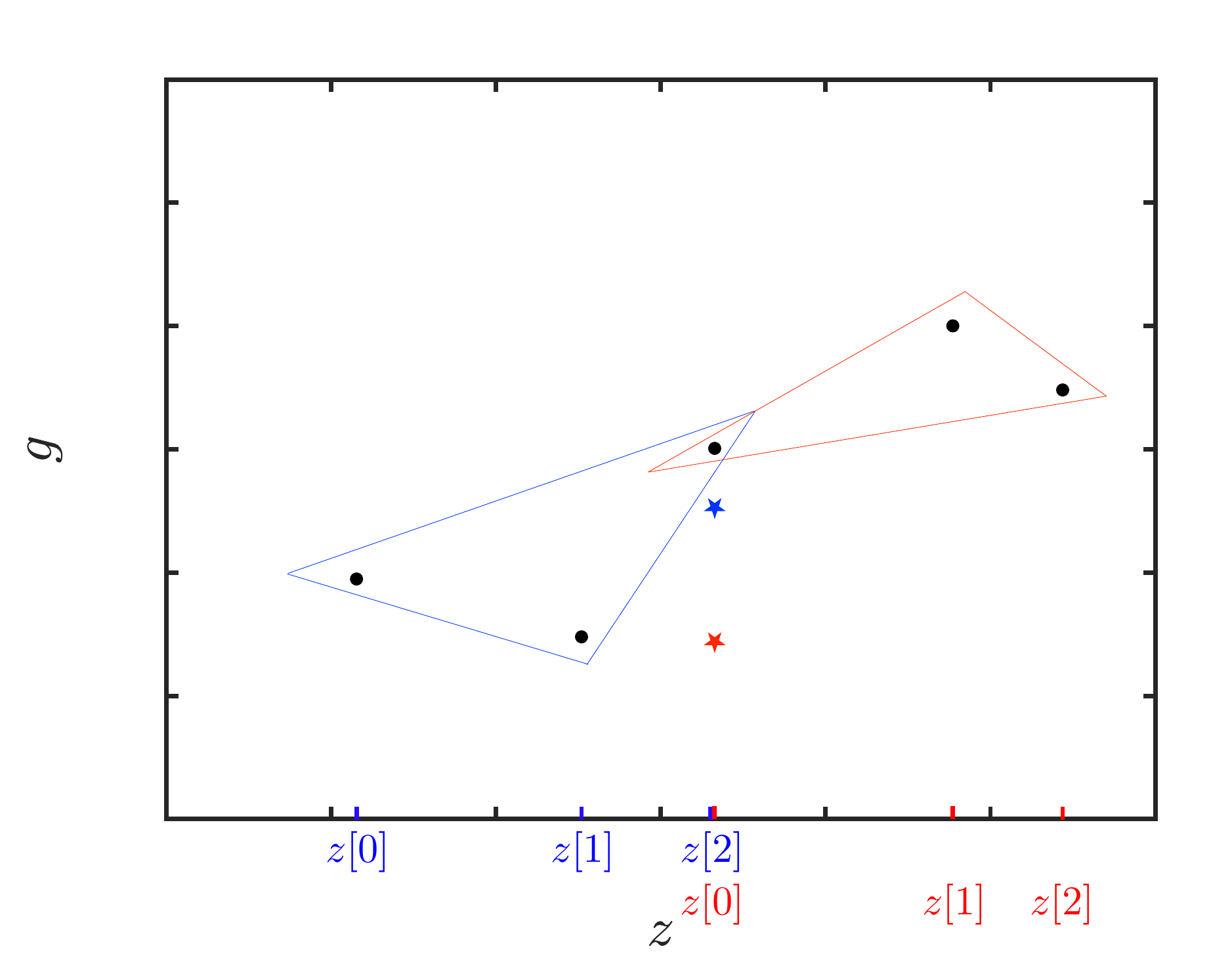}
\caption{}
\label{f:3a}
\end{center}
\end{subfigure}
\begin{subfigure}{0.28\textwidth}
\begin{center}
\includegraphics[width=\textwidth]{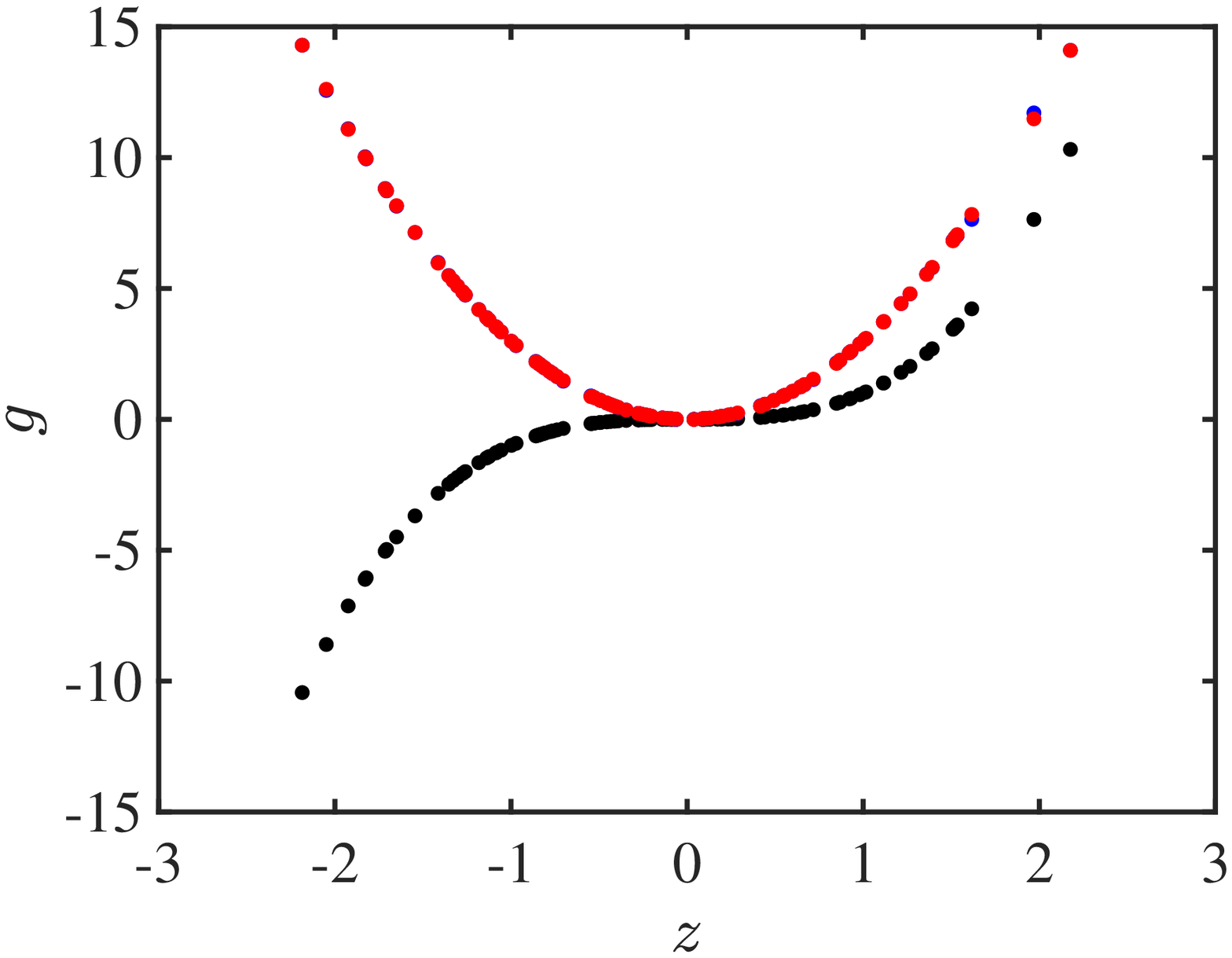}
\caption{}
\label{f:3b}
\end{center}
\end{subfigure}
\begin{subfigure}{0.28\textwidth}
\begin{center}
\includegraphics[width=\textwidth]{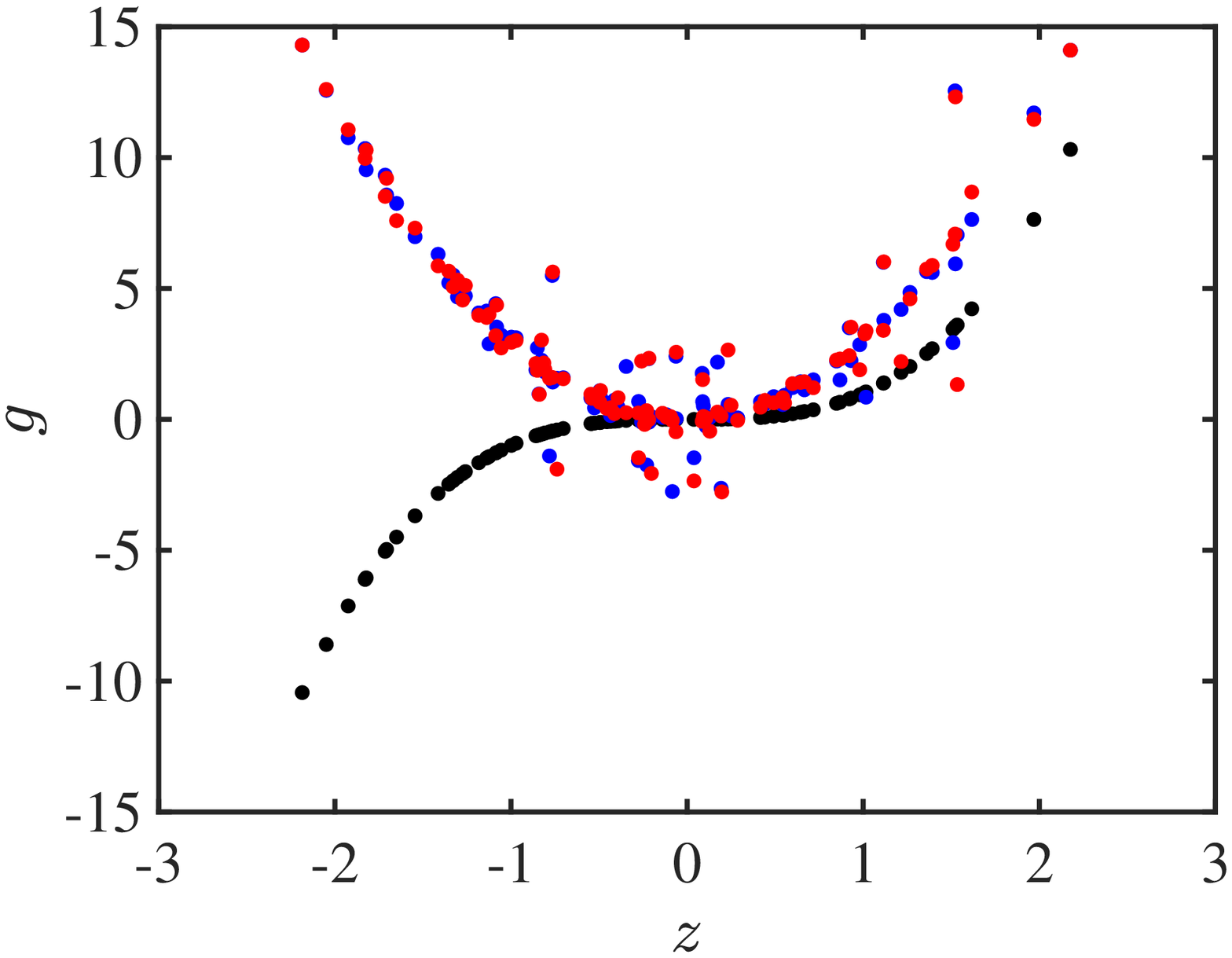}
\caption{}
\label{f:3c}
\end{center}
\end{subfigure}
\caption{(a) Black markers indicate a set of non-smooth evaluations $g(z)$. The blue and red $\star$ respectively denote the left and right 3-point finite difference, computed on the left and right window of points. (b) Black markers are evaluations of the smooth function $g = z^3$. Blue and red (almost on top of each other) are respectively the left and right 3-point finite differences. (c) Distorting the evaluations with Gaussian noise ($\sigma = 0.001\operatorname{rms}(\boldsymbol{g})$) results in clear differences amongst both finite difference approximations.}
\label{f:filter}
\end{center}
\end{figure*}

Fig.~\ref{f:3a} illustrates the operation of two distinct finite difference filters on a set of non-equidistantly spaced points. Both filters are created in accordance to Appendix \ref{a:1}. In blue, a \emph{3-point left finite difference} is used. The result is indicated with a blue $\star$. In red, a \emph{3-point right finite difference} is used (red $\star$). Both approximations yield a different result. This is due to local non-smoothness. Fig.~\ref{f:3b} and \ref{f:3c} demonstrate this property in an intuitive manner. Fig.~\ref{f:3b} depicts a number of evaluations of the smooth function $g=z^3$. In Fig.~\ref{f:3c}, these evaluations are distorted by Gaussian noise with $\sigma = 0.001 \operatorname{rms}(\boldsymbol{g})$. In both cases a 3-point left and a 3-point right finite difference is computed. The results are respectively indicated by blue and red markers. While for the smooth function hardly any difference may be observed, a clear difference is present for the distorted evaluations. This implies that the deviation between left and right finite difference filters can be used to measure local smoothness. This property will be translated into a smoothness objective. The collection of 3-point left finite difference filters will be denoted by $\mathcal{F}_L$, 3-point right filters will be denoted by $\mathcal{F}_R$ and 3-point central filters will be denoted by $\mathcal{F}_C$.

We will introduce an alternating-least-squares (ALS) algorithm in order to solve the decomposition
\begin{equation}
\mathcal{J} \approx \llbracket \boldsymbol{W}, \boldsymbol{V}, \mathcal{F}_C(\boldsymbol{V}) \circ \boldsymbol{G} \rrbracket  
\end{equation}
The matricisations of the third order tensor along its rows, columns and tubes will be denoted by $\boldsymbol{J}_{(1)}, \boldsymbol{J}_{(2)}, \boldsymbol{J}_{(3)}$, respectively, and `$\kr$' will denote the Khatri-Rao product. We may then write \citep{kolda2009}
\begin{subequations}
\begin{align}
\boldsymbol{J}_{(1)} &\approx \boldsymbol{W}((\mathcal{F}_C(\boldsymbol{V}) \circ \boldsymbol{G}) \kr \boldsymbol{V})^{\text{T}} \label{e:15a} \\ 
\boldsymbol{J}_{(2)} &\approx \boldsymbol{V}((\mathcal{F}_C(\boldsymbol{V}) \circ \boldsymbol{G}) \kr \boldsymbol{W})^{\text{T}} \label{e:15b}\\
\boldsymbol{J}_{(3)} &\approx (\mathcal{F}_C(\boldsymbol{V}) \circ \boldsymbol{G})(\boldsymbol{V} \kr \boldsymbol{W})^{\text{T}} \label{e:15c}
\end{align}
\end{subequations}
In the first step of the procedure the factors $\{\boldsymbol{W}, \boldsymbol{V},\boldsymbol{G}\}$ are initialised with random numbers. Following the ALS approach, each factor is then updated while treating both others as constants. This leads to the following objective functions and update formulas:
\begin{enumerate}
\item Update $\boldsymbol{W} \rightarrow \boldsymbol{W}^+$

From Eq.\eqref{e:15a} we infer the following objective
\begin{equation}
\underset{\boldsymbol{W}}{\operatorname{arg~min}}~\Vert \boldsymbol{J}_{(1)} - \boldsymbol{W}((\mathcal{F}_C(\boldsymbol{V}) \circ \boldsymbol{G}) \kr \boldsymbol{V})^{\text{T}} \Vert_F^2
\end{equation}
For convenience we will denote ${\boldsymbol{H}_C \coloneqq \mathcal{F}_C(\boldsymbol{V}) \circ \boldsymbol{G}}$. Given that $\boldsymbol{W}$ appears linearly in the objective, an analytical update formula is obtained,
\begin{equation}
\boldsymbol{W}^+ = \boldsymbol{J}_{(1)} \left(\left( \boldsymbol{H}_C \kr \boldsymbol{V} \right)^{\text{T}}\right)^{\dagger}
\end{equation}
Using the property
 \begin{equation}
 (\boldsymbol{A} \kr \boldsymbol{B})^{\text{T}}(\boldsymbol{A} \kr \boldsymbol{B}) = \left(\boldsymbol{A}^{\text{T}}\boldsymbol{A}\right) * \left(\boldsymbol{B}^{\text{T}}\boldsymbol{B}\right)
 \end{equation}
with `$*$' denoting the Hadamard product, leads to the more efficient update formula, requiring only the inversion of an $r \times r$ matrix
\begin{equation}
\boldsymbol{W}^+ = \boldsymbol{J}_{(1)} \left(\boldsymbol{H}_C \kr \boldsymbol{V} \right)\left(\left(\boldsymbol{H}_C^{\text{T}}\boldsymbol{H}_C\right) * \left(\boldsymbol{V}^{\text{T}} \boldsymbol{V}\right) \right)^{\dagger}
\end{equation}
\item Update $\boldsymbol{V} \rightarrow \boldsymbol{V}^+$

Apart from the tensor approximation objective, which is implied by \eqref{e:15b}, the update of $\boldsymbol{V}$ must also take into account the smoothness of $\boldsymbol{G}$. The smoothness is indeed affected by $\boldsymbol{V}$ since $\boldsymbol{V}$ defines the $z$-axis, $\boldsymbol{z} = \boldsymbol{V}^{\text{T}}\boldsymbol{p}$ (see Fig.~\ref{f:2}). The smoothness objective is therefore explicitly added using regularisation,
\begin{equation}
\label{e:V}
\begin{split}
\underset{\boldsymbol{V}}{\operatorname{arg~min}}&~\Vert \boldsymbol{J}_{(2)} - \boldsymbol{V}((\mathcal{F}_C(\boldsymbol{V}) \circ \boldsymbol{G}) \kr \boldsymbol{W})^{\text{T}} \Vert_F^2 \\
&+ \lambda \Vert \left(\mathcal{F}_L(\boldsymbol{V}) \circ \boldsymbol{G} \right) - \left(\mathcal{F}_R(\boldsymbol{V}) \circ \boldsymbol{G} \right) \Vert_F^2
\end{split}
\end{equation}
where $\lambda$ is a hyperparameter which balances both objectives. The additional term penalises divergent results from a left and a right finite difference filtering operation, ultimately steering the optimisation towards smooth solutions\footnote{In order to promote smoothness equally on all the columns of $\boldsymbol{G}$, a normalisation is required. Denoting ${\boldsymbol{H}_L \coloneqq \mathcal{F}_L \circ \boldsymbol{G}}$, and ${\boldsymbol{H}_R \coloneqq \mathcal{F}_R \circ \boldsymbol{G}}$, all columns are normalised by their rms value, i.e.\ $\boldsymbol{h}_{L_i}/\text{rms}(\boldsymbol{h}_{L_i})$, and $\boldsymbol{h}_{R_i}/\text{rms}(\boldsymbol{h}_{R_i})$.}. Given that $\boldsymbol{V}$ appears nonlinearly in \eqref{e:V}, nonlinear optimisation is required to obtain $\boldsymbol{V}^+$.

\item Update $\boldsymbol{G} \rightarrow \boldsymbol{G}^+$

In analogy with \eqref{e:V}, the update formula of $\boldsymbol{G}$ is also found from a joint objective function.
\begin{equation}
\label{e:G}
\begin{split}
\underset{\boldsymbol{G}}{\operatorname{arg~min}}&~\Vert \boldsymbol{J}_{(3)} -(\mathcal{F}_C(\boldsymbol{V}) \circ \boldsymbol{G})(\boldsymbol{V} \kr \boldsymbol{W})^{\text{T}} \Vert_F^2 \\
&+ \lambda \Vert \left(\mathcal{F}_L(\boldsymbol{V}) \circ \boldsymbol{G} \right) - \left(\mathcal{F}_R(\boldsymbol{V}) \circ \boldsymbol{G} \right) \Vert_F^2
\end{split}
\end{equation}
It can be shown that $\boldsymbol{G}$ appears linearly in \eqref{e:G}. This allows for an analytical update formula (to be consulted in Appendix \ref{a:B1}).
\end{enumerate}

Steps (1) to (3) are iterated until convergence is reached or a stopping criteria is met. In each step the updated factors are passed on to the next. The appropriate value of $\lambda$ is to be determined from a line search.
The univariate functions $g_i(z_i)$ are ultimately obtained from parameterising the columns $\boldsymbol{g}_i$ using a basis expansion which can be freely chosen by the user.

\section{Toy problem}
\label{s:toy}

In this section, the Filtered CPD method is demonstrated on a toy problem. The example will be of the polynomial class, although it should be stressed that this is not a requirement. 

Consider the decoupled polynomial function
\begin{equation}\label{e:dec}
\boldsymbol{f}(\boldsymbol{p}) = \overbrace{\left[ \begin{array}{ccc}  3 & 0.5 & -1 \\ 1& 2 & 3 \end{array} \right]}^{\boldsymbol{W}} \overbrace{\left[ \begin{array}{c}  z_1^3 + 0.5 z_1^2 \\ 2 z_2^3 + z_2^2 \\ z_3^3 + 3 z_3^2 \end{array} \right]}^{\boldsymbol{g}},~
\boldsymbol{z}= \overbrace{\left[ \begin{array}{cc}  1 & 2 \\ 3 & 1 \\ 0.5 & 3 \end{array}\right]}^{\boldsymbol{V}^{\text{T}}} \boldsymbol{p}
\end{equation} 
for which $m=2$, $n=2$ and $r=3$. An equivalent formulation\footnote{The equivalency is subject to rounding errors in the computation of the coefficients.}, expressed in the standard monomial basis reads
\begin{equation}\label{e:coupled}
\resizebox{0.432\textwidth}{!}{$
\boldsymbol{f}(\boldsymbol{p}) = \left[ \begin{array}{ccccccc}  5.25 & 0 & -20.5 & 29.875 & 42.75 & 31.5 & -2 \\ 20.75 & 41 & 85 & 109.375 & 120.75 & 88.5 & 93 \end{array} \right] \begin{bmatrix} p_1^2 \\ p_1p_2 \\ p_2^2 \\ p_1^3 \\ p_1^2 p_2 \\ p_1 p_2^2 \\ p_2^3 \end{bmatrix}
$}
\end{equation}
One can think of \eqref{e:coupled} as a static multivariate polynomial although it might as well be the polynomial which describes a NARX model. In that case $p_1$ and $p_2$ would represent (delayed) input or output samples and $n$ would be equal to 1. Furthermore, it could be the nonlinear function which describes the state update in nonlinear state-space models, in which case $p_1$ and $p_2$ would be either state or input variables \citep{decuyper2021}.

The objective is to retrieve the decoupled function \eqref{e:dec} starting from \eqref{e:coupled}. 
The first step consists in evaluating the Jacobian of \eqref{e:coupled}, $\boldsymbol{J}$, and stacking the matrices in the third dimension. A number of $N=100$ operating points were randomly selected \rev{from $\mathcal{U}\left(-1.5,1.5\right)$}, leading to $\mathcal{J} \in \mathbb{R}^{2 \times 2 \times 100}$. In practice the operating points should cover the domain of interest.

The next step involves decomposing the Jacobian tensor. At this stage the number of univariate branches, allowed in the decoupled function, must be determined. For the sake of illustration $r=3$ is selected, for we know that an exact decoupled function of the form exists \eqref{e:dec}. In practice a scan over $r$ may be required.

\textbf{CPD}:
First we examine the classical canonical polyadic decomposition (CPD), which returns
 ${\mathcal{J} \approx \llbracket \boldsymbol{W}, \boldsymbol{V}, \boldsymbol{H} \rrbracket}$. With a tensor approximation \eqref{e:5} close to machine precision, the decomposition was successful. Studying the result, however, reveals an underlying problem (depicted in Fig.~\ref{f:2}). As may be observed, the decomposition did not convey information on the univariate functions, which make up \eqref{e:dec}. The decomposition only points to \eqref{e:dec} in case it is unique, which is not the case in this example for $r=3$.
 
\begin{figure}
\begin{center}
\includegraphics[width=0.4\textwidth]{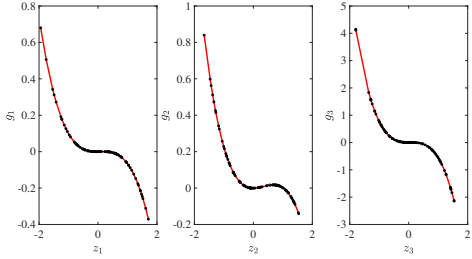}
\caption{Factor $\boldsymbol{G}$ obtained from a filtered CPD using an explicit smoothness objective with $\lambda = 10^2$. Black markers represent the elements of the columns of $\boldsymbol{G}$, red corresponds to a polynomial fit.}
\label{f:4}
\end{center}
\end{figure}

 \textbf{F-CPD}: Not all solutions of the decomposition prove useful in our search for the decoupled function. We can, however, steer the decomposition towards appropriate solutions by promoting smoothness. The result of explicitly adding a smoothness objective is depicted in Fig.~\ref{f:4}. It was selected from a coarse search over $\lambda = 10^{-3}, 10^{-2},10^{-1},1,10,10^2,10^3$, eventually settling for $\lambda = 10^2$. The black markers clearly describe smooth functions. Balancing both objectives results in a tensor approximation with a relative error of 0.7\%.
 
To finalise the decoupling procedure, a parametric fit is obtained from $\boldsymbol{G}$. In Fig.~\ref{f:4} an accurate third order polynomial is fitted (red lines). To assess the accuracy of the decoupled function, a relative root-mean-squared error value is computed. Denoting the decoupled function $\boldsymbol{f}_d(\boldsymbol{p}) = \boldsymbol{W} \boldsymbol{g}\left(\boldsymbol{V}^{\text{T}}\boldsymbol{p}\right)$, we compute an error metric for every entry of $\boldsymbol{f}_d$,
 \begin{equation}
 \label{e:rel_er}
 e_i = \frac{\sqrt{\frac{1}{N}\sum_{k=1}^N\left(f_i(\boldsymbol{p}[k]-f_{d_i}(\boldsymbol{p}[k])\right)^2}}{\sqrt{\frac{1}{N}\sum_{k=1}^N f_i(\boldsymbol{p}[k])^2}} \times 100
 \end{equation}
The decoupled function is an accurate approximation of \eqref{e:coupled}, yielding relative errors $e_1 = 0.6$\% and $e_2 =0.6 $\%. The exact decoupled function, \eqref{e:dec} is, however, not retrieved given that the ALS algorithms lead to local minima of the objectives. 


\section{Conclusion}
Decoupling multivariate functions is a useful model reduction technique. It reveals underlying structure while potentially reducing the number of parameters. An established procedure is to compute a collection of Jacobian matrices, store them in a three-way array, and resort to a diagonal tensor decomposition to retrieve the decoupled function. In practice, however, the tensor decomposition tends to be non-unique, preventing the retrieval of an appropriate solution. In this work, non-parametric finite difference filters are used to promote smooth solutions. Smoothness ensures that the decomposition points to a decoupled function. A toy problem illustrates that accurate decoupled forms can be retrieved.

\bibliography{entirebib}             

\begin{thebibliography}{8}
\providecommand{\natexlab}[1]{#1}
\providecommand{\url}[1]{\texttt{#1}}
\providecommand{\urlprefix}{URL }
\expandafter\ifx\csname urlstyle\endcsname\relax
  \providecommand{\doi}[1]{doi:\discretionary{}{}{}#1}\else
  \providecommand{\doi}{doi:\discretionary{}{}{}\begingroup
  \urlstyle{rm}\Url}\fi

\bibitem[{Billings(2013)}]{billings2013}
Billings, S. (2013).
\newblock \emph{Nonlinear System Identification: {NARMAX} Methods in the Time,
  Frequency and Spatio-Temporal Domains}.
\newblock Wiley.

\bibitem[{Decuyper et~al.(2019)Decuyper, Dreesen, Schoukens, Runacres, and
  Tiels}]{decuyper2019}
Decuyper, J., Dreesen, P., Schoukens, J., Runacres, M., and Tiels, K. (2019).
\newblock Decoupling multivariate polynomials for nonlinear state-space models.
\newblock \emph{{IEEE} {C}ontrol {S}ystems {L}etters}, 3(3), 745--750.

\bibitem[{Decuyper et~al.(2021)Decuyper, Tiels, Runacres, and
  Schoukens}]{decuyper2021}
Decuyper, J., Tiels, K., Runacres, M.C., and Schoukens, J. (2021).
\newblock Retrieving highly structured models starting from black-box nonlinear
  state-space models using polynomial decoupling.
\newblock \emph{{Mechanical Systems And Signal Processing}}, 146.

\bibitem[{Dreesen et~al.(2018)Dreesen, De~Geeter, and Ishteva}]{dreesen2018}
Dreesen, P., De~Geeter, J., and Ishteva, M. (2018).
\newblock Decoupling multivariate functions using second-order information and
  tensors.
\newblock In \emph{Proc. 14th International Conference on Latent Variable
  Analysis and Signal Separation {LVA/ICA} 2018}, volume 10891, 79--88.
  Springer.

\bibitem[{Dreesen et~al.(2015)Dreesen, Ishteva, and Schoukens}]{dreesen2014}
Dreesen, P., Ishteva, M., and Schoukens, J. (2015).
\newblock Decoupling multivariate polynomials using first-order information.
\newblock \emph{{SIAM} Journal on Matrix Analysis and Applications}, 36(2),
  864--879.

\bibitem[{Karami et~al.(2021)Karami, Westwick, and Schoukens}]{karami2021}
Karami, K., Westwick, D., and Schoukens, J. (2021).
\newblock Applying polynomial decoupling methods to the polynomial {NARX}
  model.
\newblock \emph{{Mechanical Systems And Signal Processing}}, 148.

\bibitem[{Kolda and Bader(2009)}]{kolda2009}
Kolda, T.G. and Bader, B.W. (2009).
\newblock Tensor decomposition and applications.
\newblock \emph{{SIAM} {R}ev.}, 51(3), 455--500.

\bibitem[{Singh and Bhadauri(2009)}]{singh2009}
Singh, A.K. and Bhadauri, B.S. (2009).
\newblock Finite difference formulae for unequal subintervals using
  {Lagrange's} interpolation formula.
\newblock \emph{{Int. Journal of Math. Analysis}}, 3(17), 815--827.

\end{thebibliography}
                                                   







\appendix
\vspace{-0.04cm}
\section{Non-equidistant finite differences}    
\label{a:1}
\rev{A finite difference expression for non-equidistantly spaced points can be obtained through Lagrange interpolation \citep{singh2009}. Consider $\boldsymbol{z}_w$ to be a window of $k$ neighbouring elements of $z$ and let $\boldsymbol{g}_w$ be the corresponding window of function evaluations. The Lagrange polynomial is the polynomial of lowest degree which interpolates all $k$ points within the window. It is a linear combination of basis functions
\begin{equation}
L(z) \coloneqq \sum_{j=1}^k \boldsymbol{g}_w[j]l_j(z)
\end{equation}
where $l_j$ are known as Lagrange basis polynomials
\begin{equation}
l_j(z) \coloneqq \prod_{\begin{matrix} 1 \le i \le k \\ i \ne j \end{matrix}} \frac{z - \boldsymbol{z}_w[i]}{\boldsymbol{z}_w[j]-\boldsymbol{z}_w[i]}
\end{equation}
At the grid points $\boldsymbol{z}_w$, the basis polynomials have the following two properties:
\begin{enumerate}
\item For all $s \ne j$ the product will be zero since the numerator $\boldsymbol{z}_w[s] - \boldsymbol{z}_w[i]$ attains zero at $s = i$. Hence $\forall (s \ne j) :$
\begin{equation}
l_j(\boldsymbol{z}_w[s]) = \prod_{\begin{matrix} 1 \le i \le k \\ i \ne j \end{matrix}} \frac{\boldsymbol{z}_w[s] - \boldsymbol{z}_w[i]}{\boldsymbol{z}_w[j]-\boldsymbol{z}_w[i]} = 0 
\end{equation}
\item On the other hand
\begin{equation}
l_j(\boldsymbol{z}_w[j]) = \prod_{\begin{matrix} 1 \le i \le k \\ i \ne j \end{matrix}} \frac{\boldsymbol{z}_w[j] - \boldsymbol{z}_w[i]}{\boldsymbol{z}_w[j]-\boldsymbol{z}_w[i]} = 1
\end{equation}
\end{enumerate}
Given that all basis polynomials are zero when ${s \ne j}$, and since they equal to one for $s =j$, it follows that $L(\boldsymbol{z}_w[j]) = \boldsymbol{g}_w[j]$, meaning that $L(z)$ passes exactly through all $k$ points in the window. Having found a function which relates neighbouring points, also a derivative relationship between neighbouring points (i.e.\ a finite difference) may be obtained. A first order derivative expression is given by
\begin{equation}
\label{e:L1}
L^{(1)}(z) \coloneqq \sum_{j=1}^k \boldsymbol{g}_w[j]l^{(1)}_j(z) 
\end{equation}
which is a weighted sum with the weights following from $l^{(1)}_j(z) \coloneqq \frac{dl_j(z)}{dz}$
\begin{equation}
\resizebox{0.42\textwidth}{!}{$
l^{(1)}_j(z) \coloneqq \sum_{\begin{matrix} s = 1 \\ s \ne j \end{matrix}}^k \left[ \frac{1}{\boldsymbol{z}_w[j] - \boldsymbol{z}_w[s]} \prod_{\begin{matrix} i =1 \\ i \ne (s,j) \end{matrix}}^k \frac{z - \boldsymbol{z}_w[i]}{\boldsymbol{z}_w[j] - \boldsymbol{z}_w[i]}  \right]$}
\end{equation}


Notice that the linear combination, expressed by \eqref{e:L1}, corresponds to the operations encoded in the rows of the matrix $\boldsymbol{D}$ \eqref{e:D}. When computing a 3-points finite difference, $k=3$, the weights $l^{(1)}_j(z)$ with $j=1,\ldots,k$ form the elements of $\boldsymbol{D}$. Contrary to \eqref{e:D}, all weights may differ from one another since they are computed on the basis of the non-equidistant spacing of the $k$ points in $\boldsymbol{z}_w$. Going from one row to the next may be seen as sliding the window of the finite difference operator over the $z$-axis. The positioning of the window allows switching between forward, central, and backward differencing, additionally also the length may be changed. This flexibility enables to compute a number of alternative finite difference filters, which are all approximations of the first order derivative, e.g.\ a \emph{{3-points central finite difference filter}}. The $i$th row of a 3-points ($k=3$) central differencing filter of size $N$, denoted $\boldsymbol{D}_C$, is based on the window $\boldsymbol{z}_{w_i} = \left[ \boldsymbol{z}_s[i-1] \quad \boldsymbol{z}_s[i] \quad \boldsymbol{z}_s[i+1] \right]^{\text{T}}$, with $\boldsymbol{z}_s$ the sorted version of $\boldsymbol{z}$. The weights are accordingly found by evaluating the basis functions $l_j^{(1)}(z)$ in the central point of the window, i.e.\ $l_j^{(1)}(\boldsymbol{z}_{w_i}[2])$ with $j=1,\ldots,k$. The following filter is obtained:

\renewcommand{\arraystretch}{3}

\begin{equation}
\label{e:D1}
\resizebox{0.4\textwidth}{!}{$
\boldsymbol{D}_C \coloneqq  \left[ \begin{array}{cccccccc}
 l_1^{(1)}(\boldsymbol{z}_{w_1}[1]) &  l_2^{(1)}(\boldsymbol{z}_{w_1}[1]) & l_3^{(1)}(\boldsymbol{z}_{w_1}[1]) & 0 & \dots &  0 \\
  l_1^{(1)}(\boldsymbol{z}_{w_2}[2]) &  l_2^{(1)}(\boldsymbol{z}_{w_2}[2]) & l_3^{(1)}(\boldsymbol{z}_{w_2}[2]) & 0 & \dots &  0 \\
   0 &  l_1^{(1)}(\boldsymbol{z}_{w_3}[2]) &  l_2^{(1)}(\boldsymbol{z}_{w_3}[2]) & l_3^{(1)}(\boldsymbol{z}_{w_3}[2]) & \ddots & \vdots \\
 \vdots & \ddots & \ddots & \ddots & \ddots & 0  \\
  0 & \dots & 0 & l_1^{(1)}(\boldsymbol{z}_{w_{N-1}}[2]) &  l_2^{(1)}(\boldsymbol{z}_{w_{N-1}}[2]) & l_3^{(1)}(\boldsymbol{z}_{w_{N-1}}[2])\\
    0 & \dots & 0 & l_1^{(1)}(\boldsymbol{z}_{w_{N}}[3]) &  l_2^{(1)}(\boldsymbol{z}_{w_{N}}[3]) & l_3^{(1)}(\boldsymbol{z}_{w_{N}}[3])\\
 \end{array} \right] $}
\end{equation}
The first and the last row face a boundary and are treated as 3-points right (backward) and 3-points left (forward) filters, correspondingly. In an analogue way also a left and a right derivative approximation may be constructed by shifting the window left or right of the diagonal. }

\section{Update formula of $\boldsymbol{G}$}
\label{a:B}

\label{a:B1}
 \emph{Lemma 1.\ } Let \eqref{e:G} be the objective function from which $\boldsymbol{G}$ is to be updated. The claim is that $\boldsymbol{G}$ appears linearly in this joint objective such that an analytical update formula may be derived. 

\begin{equation} \tag{\ref{e:G}}
\begin{split}
\underset{\boldsymbol{G}}{\operatorname{arg~min}}&~\left\| \boldsymbol{J}_{(3)} -(\mathcal{F}_C(\boldsymbol{V}) \circ \boldsymbol{G})(\boldsymbol{V} \kr \boldsymbol{W})^{\text{T}} \right\|_F^2 \\
&+ \lambda \Vert \left(\mathcal{F}_L(\boldsymbol{V}) \circ \boldsymbol{G} \right) - \left(\mathcal{F}_R(\boldsymbol{V}) \circ \boldsymbol{G} \right) \Vert_F^2 
\end{split}
\end{equation}

 \emph{Proof: } Consider the vectorised form of the objective function. For clarity the dependence on $\boldsymbol{V}$ is dropped, i.e.\ $\mathcal{F}_C(\boldsymbol{V}) = \mathcal{F}_C$. 

\begin{equation} \label{e:obj_vec}
\resizebox{0.42\textwidth}{!}{$
\begin{split}
&\underset{\boldsymbol{G}}{\operatorname{arg~min}}~\left\| \text{vec}(\boldsymbol{J}_{(3)}) -  \operatorname{vec}\left(\left(\mathcal{F}_C \circ \boldsymbol{G}\right)\left(\boldsymbol{V} \kr \boldsymbol{W} \right)^{\text{T}} \right) \right\|_F^2 \\
&+\lambda \Vert  \text{vec}\left(\mathcal{F}_L \circ \boldsymbol{G} \right)  -\text{vec} \left(\mathcal{F}_R \circ \boldsymbol{G}\right) \Vert_F^2
\end{split}
$}
\end{equation}

 Denoting the Knonecker product by `$\kron$' we may use the property, $\text{vec}\left(\boldsymbol{A}\boldsymbol{X}\boldsymbol{B}\right) = (\boldsymbol{B}^{\text{T}} \kron \boldsymbol{A})\text{vec}(\boldsymbol{X})$. With $\boldsymbol{A} = \boldsymbol{I}_N$, $\boldsymbol{B} = \left(\boldsymbol{V} \kr \boldsymbol{W} \right)^{\text{T}}$ and $\boldsymbol{X} = \left(\mathcal{F}_C \circ \boldsymbol{G} \right)$ we have that 

\begin{equation}
\label{e:AXB}
\begin{split}
\operatorname{vec}&\left(\left(\mathcal{F}_C \circ \boldsymbol{G}\right) \left(\boldsymbol{V} \kr \boldsymbol{W} \right)^{\text{T}} \right) =\\ 
&\quad \quad \quad \left( \left(\boldsymbol{V} \kr \boldsymbol{W}\right) \kron \boldsymbol{I}_N \right) \text{vec}\left(\mathcal{F}_C \circ \boldsymbol{G} \right)
\end{split}
\end{equation}
 Substituting \eqref{e:AXB} and applying the property of \eqref{e:vec}, the objective is rewritten into 
\begin{equation}
\resizebox{0.42\textwidth}{!}{$
\begin{split}
&\underset{\boldsymbol{G}}{\operatorname{arg~min}}~\left\| \text{vec}(\boldsymbol{J}_{(3)}) \right. \\
& -\left. \left( \left(\boldsymbol{V} \kr \boldsymbol{W}\right) \kron \boldsymbol{I}_N \right) \text{blkdiag}\left(\mathcal{F}_C\right) \text{vec}\left(\boldsymbol{G}\right) \right\|_F^2 \\
&+\lambda \Vert \left(\text{blkdiag}\left(\mathcal{F}_L\right) -\text{blkdiag} \left(\mathcal{F}_R \right) \right)\text{vec}(\boldsymbol{G}) \Vert_F^2
\end{split}
$}
\end{equation}
Reordering both terms leads to the expression 
\begin{equation}
\begin{split}
&\underset{\boldsymbol{G}}{\operatorname{arg~min}}~\left\| \begin{bmatrix} \text{vec}(\boldsymbol{J}_{(3)}) \\ \boldsymbol{0} \end{bmatrix} \right. \\
&- \left.  \begin{bmatrix}  \left( \left(\boldsymbol{V} \kr \boldsymbol{W}\right) \kron \boldsymbol{I}_N \right) \text{blkdiag}\left(\mathcal{F}_C \right) \\ \lambda  \left(\text{blkdiag}\left(\mathcal{F}_L\right) -\text{blkdiag} \left(\mathcal{F}_R \right) \right) \end{bmatrix} \text{vec}(\boldsymbol{G}) \right\|_F^2
\end{split}
\end{equation}
where $\boldsymbol{0} \in \mathbb{R}^{rN}$ is a vector of zeros and from which it is clear that $\text{vec}(\boldsymbol{G})$ enters linearly in the objective. \qed 

An analytical update formula is then obtained by defining 
\begin{equation}
\boldsymbol{K} \coloneqq  \begin{bmatrix}  \left( \left(\boldsymbol{V} \kr \boldsymbol{W}\right) \kron \boldsymbol{I}_N \right) \text{blkdiag}\left(\mathcal{F}_C \right) \\ \lambda  \left(\text{blkdiag}\left(\mathcal{F}_L\right) -\text{blkdiag} \left(\mathcal{F}_R \right) \right) \end{bmatrix},
\end{equation}

 leading to the least-squares solution 

\rev{
\begin{equation}
\text{vec}(\boldsymbol{G}^+) = \boldsymbol{K}^{\dagger} \begin{bmatrix} \text{vec}(\boldsymbol{J}_{(3)}) \\ \boldsymbol{0} \end{bmatrix}
\end{equation}
}

 To ensure that smoothness is promoted equally on all columns of $\boldsymbol{G}$, a normalisation is required (similar to the update of $\boldsymbol{V}$). The block diagonal filters are therefore scaled by the root-mean-squared value of their corresponding columns. 

\begin{equation}
\resizebox{0.4\textwidth}{!}{$
\label{e:vec2}
\text{blkdiag}(\mathcal{F}_L) = \left[\begin{array}{cccc}\frac{\boldsymbol{F}_{L_1}}{\text{rms}(\boldsymbol{h}_{L_1})} & 0 & \dots & 0 \\ 0 & \frac{\boldsymbol{F}_{L_2}}{\text{rms}(\boldsymbol{h}_{L_2})} & \ddots & \vdots \\ \vdots & \ddots & \ddots & 0 \\ 0 & \cdots & 0 & \frac{\boldsymbol{F}_{L_r}}{\text{rms}(\boldsymbol{h}_{L_r})} \end{array} \right]
$}
\end{equation}
In an analogue way $\text{blkdiag}(\mathcal{F}_R)$ is normalised. 
\end{document}